# THE INTERFERENCE TERM IN THE WIGNER DISTRIBUTION FUNCTION AND THE AHARONOV-BOHM EFFECT

D. Dragoman[*] – Univ. Bucharest, Physics Dept., P.O. Box MG-11, 76900 Bucharest, Romania

ABSTRACT

A phase space representation of the Aharonov-Bohm effect is presented. It shows that the shift of interference fringes is associated to the interference term of the Wigner distribution function of the total wavefunction, whereas the interference pattern is defined by the common projections of the Wigner distribution functions of the interfering beams.

[*] Correspondence address: D. Dragoman, P.O. Box 1-480, 70700 Bucharest, Romania, tel./fax: +40-1-6473382, email: danieladragoman@yahoo.com

The continuous struggle for an intuitive interpretation of quantum mechanics has been significantly supported by the phase space interpretation of quantum mechanics, which does not involve operators, but only numerical variables. In particular, the Wigner distribution function (WDF) [1,2] was used to understand the difference between the situations when quantum mechanical wavefunctions interfere with one another or when transitions between quantum states are to be expected. More specifically, it was shown that quantum transitions between different quantum states occur whenever the WDF of the two states overlap [3-5], whereas interferences occur when the WDF of the interfering states do not overlap, but have non-zero values along common intervals of the $x$ and/or $p$ coordinates [6,7]. The aim of the present paper is to describe the Aharonov-Bohm effect in phase space. Such an approach towards the explanation of this effect will reveal the subtleties of the phase space interpretation of quantum interference, and will contribute to a better understanding of the role played by different terms of the WDF of a superposition of states.

To begin with, we consider the two-slit quantum interference, in which, for simplicity, we assume that the two wavefunctions associated to the two slits/paths in the coordinate representations are Gaussians, such that the total wavefunction of the system is given by $\varphi(x) = \exp[-(x-d)^2/2x_0^2] + \exp[-(x+d)^2/2x_0^2]$. If the distance $2d$ between the two slits is chosen sufficiently large, immediately after the slits the two wavefunctions are spatially separated, such that interference can only occur after propagation over a certain distance $D$. The WDF of the (pure) quantum system described by the wavefunction above is defined as

$$W(x,p) = \int \varphi^*(x-x'/2)\varphi(x+x'/2)\exp(ipx'/\hbar)dx' \qquad (1)$$

and is given by

$$W(x,p) = 2x_0\pi^{1/2}\exp(-p^2x_0^2/\hbar^2)\{\exp[-(x-d)^2/x_0^2] + \exp[-(x+d)^2/x_0^2] + 2\exp(-x^2/x_0^2)\cos(2pd/\hbar)\} \qquad (2)$$

The first two terms in equation (2) correspond to the WDFs of the individual slits, whereas the last term, with an oscillatory behavior is the interference term. In Ref.6 it was argued that the WDFs of the individual slits are a measure of the phase space localizability of the quantum particles that pass through these slits. The contour plot of the WDF in (2) is shown in Fig.1a, where the darker regions correspond to lower values of the WDF. The WDFs of the individual slits are seen as the two outer Gaussian terms (the distance *d* was chosen equal to $5x_0$) with positive values, whereas the inner interference term shows an oscillatory behavior. The normalized coordinates in Fig.1a (as in the rest of the figures) are defined as $X = x/x_0$, $P = px_0/\hbar$. In a previous paper [6] it was shown that interferences in the *x* and/or *p* domains occur only when the WDFs of the individual wavefunctions have non-vanishing values along common intervals of the corresponding variables. The interferences in the *x* and *p* domains manifest themselves as oscillatory behaviors of $|\varphi(x)|^2$ and $|\bar{\varphi}(p)|^2$ respectively, where

$$\bar{\varphi}(p) = \int \varphi(x)\exp(ixp/\hbar)dx \tag{3}$$

is the Fourier transform of the wavefunction associated to the quantum system in the coordinate representation. $|\varphi(x)|^2$ and $|\bar{\varphi}(p)|^2$, represented in Figs.1b and 1c, respectively, are related to the WDF through

$$|\varphi(x)|^2 = 1/(2\pi\hbar)\int W(x,p)dp, \quad |\bar{\varphi}(p)|^2 = \int W(x,p)dx \tag{4}$$

Since the WDFs of the individual slits have a common interval of non-vanishing values along the *p* direction and are well separated along *x*, interference occurs only along the *p* direction. Interferences along *p*, in agreement with a phase space interpretation of quantum interference, have been experimentally observed in [8]. Note that the maxima in $|\bar{\varphi}(p)|^2$ correspond to the maxima in the oscillatory interference term of the WDF. Interference along the *x* direction can only be observed after the quantum particles propagate a sufficient long distance such that the WDFs of the particles passing through the two slits have a common interval of projection along *x*. The WDF suffers a shear transformation at propagation along a distance *D* in free space, described by $W(x,p;z=D) = W(x-\alpha p, p; z=0)$ where $\alpha = t/m$ for a quantum system consisting of particles with mass *m* which traverse the distance *D* in a time *t* [9]. The WDF, as well as $|\varphi(x)|^2$ and $|\bar{\varphi}(p)|^2$, after a distance $\alpha = 6x_0^2/\hbar$ are shown in Figs.2a, 2b and 2c respectively. Interference is now expected also along the *x* domain where the two sheared Gaussians representing the WDFs of the two slits have a common projection interval. Note that $|\bar{\varphi}(p)|^2$ remains constant at propagation in free space. A simple calculation shows that $|\bar{\varphi}(p)|^2 \approx \exp(-p^2 x_0^2/\hbar^2)\cos^2(pd/\hbar)$ at both at $z = 0$ and $z = D$.

The (magnetic) Aharonov-Bohm effect [10] consists of a shift in the interference fringes, inside the same interference pattern, when the charged quantum particles passing through the two slits are traveling along two paths on which the magnetic field **B** vanishes but the vector potential **A** defined through $\mathbf{B} = \nabla \times \mathbf{A}$ is different from zero. The parameter that characterizes the shift of the interference fringes *inside the same interference pattern* is the difference between the phases acquired by the two beams of particles passing through the two slits, given by

$$\delta = (q/\hbar c)\left(\int_{path1} \mathbf{A} \cdot d\mathbf{s} - \int_{path2} \mathbf{A} \cdot d\mathbf{s}\right) = 2\pi \Phi / \Phi_0 \qquad (5)$$

where $q$ is the charge of the particles, $\Phi$ is the magnetic flux and $\Phi_0 = hc/e$.

In order to sense the essential features of the Aharonov-Bohm effect we suppose that the two beams of particles acquire each the phases $\delta/2$ and $-\delta/2$, respectively, in the region immediately after the two slits, such that the initial total wavefunction becomes $\varphi(x) = \exp[-(x-d)^2/2x_0^2]\exp(-i\delta/2) + \exp[-(x+d)^2/2x_0^2]\exp(i\delta/2)$. The total WDF is now

$$W(x,p) = 2x_0\pi^{1/2}\exp(-p^2 x_0^2/\hbar^2)\{\exp[-(x-d)^2/x_0^2] + \exp[-(x+d)^2/x_0^2] + 2\exp(-x^2/x_0^2)\cos(2pd/\hbar - \delta)\} \qquad (6)$$

It is interesting to note that only the interference term changes, whereas the WDFs of the individual slits remain constant. Practically, the oscillations in the interference term shift upwards or downwards depending on the sign of $\delta$; the shift in the interference term of the WDF is easily observable in Fig.3a, for which $\delta = 4$. Our phase space interpretation suggests that no interference along the *x* axis can be observed immediately after the slits, but the interference fringes along the *p* axis shift with $\delta$, the interference pattern (the interval along the *p* axis with common projections of the WDFs of the individual slits) remaining the same. A simple calculation shows that now $|\overline{\varphi}(p)|^2 \approx \exp(-p^2 x_0^2/\hbar^2)\cos^2(pd/\hbar - \delta/2)$, the interference fringes along the *p* direction *shifting with the same amount* as the interference fringes which form along the *x* direction after free space propagation over a sufficiently long distance. Although such a shift in the interference fringes along the *p* direction should be expected since in the presence of the vector potential the momentum operator **p** changes to $\mathbf{p} - q\mathbf{A}/c$, we are not aware of any experimental test of this feature of the interference pattern in the presence of the vector potential. This Aharonov-Bohm-like effect in the $p$ coordinate is – until experimentally confirmed – a *prediction* of the phase space interpretation of quantum interference. The probability of finding the quantum particles after propagation over a distance $D$, and in the presence of a phase shift $\delta$ is $|\varphi(x)|^2 = (x_0/\Delta)\{\exp[-(x-d)^2/\Delta^2] + \exp[-(x+d)^2/\Delta^2] + 2\exp[-(x^2+d^2)/\Delta^2]\cos[2x\alpha d\hbar/(x_0^2\Delta^2) - \delta]\}$, where $\Delta = \sqrt{\alpha^2\hbar^2 + x_0^4}/x_0$. The change in $|\varphi(x)|^2$ due to the Aharonov-Bohm effect appears as a shift with $\delta$ in the interference fringes at large enough distances where the interfering wavefunctions overlap (Fig.4b). The shift in the interference fringes along the $x$ domain occurs inside the same interference pattern, defined as the $x$ domain where the WDFs of the two individual slits have a common projection. From the expression of the WDF in (6) it follows that the interference fringes appear unchanged for a magnetic flux integer multiple of $\Phi_0$. These results, which characterize he Aharonov-Bohm effect, appear as a natural consequence of the phase space treatment, when we correctly describe the role of the interference term of the total WDF. It is the term which characterizes the *potentiality* of interference, and also defines the position of the interference extrema in the $p$ and $x$ domains, the *actual occurrence* of the interference being however described by the WDFs of the interfering wavefunctions, more precisely by their common projections along the $x$ and $p$ axis. Moreover, the interference term in the total WDF defines the interference fringes, whereas the WDFs of the interfering wavefunctions define the interference pattern.

    It is worth mentioning that the same phase space treatment of the magnetic, or vector Aharonov-Bohm effect can be also applied to the scalar Aharonov-Bohm effect, since in the mathematical treatment no explicit mention to the origin of the phase shift is made. In the scalar Aharonov-Bohm effect with electrons the phase shift is induced not by the vector potential, but by the scalar potential in the Schrödinger equation [11]. Actually, the two interfering electron wavepackets pass through conducting cylinders that act as Faraday cages, on which voltage pulses $U(t)$ are applied, the phase shift being given by $\delta = (e/\hbar)(\int_{path1} U(t)dt - \int_{path2} U(t)dt)$. Although the scalar Aharonov-Bohm effect with electrons was not confirmed experimentally up to now due to technical difficulties, a similar effect was evidenced using neutrons [12]. In the scalar Aharonov-Bohm effect with neutrons [13] the electromagnetic field interacts with the neutron magnetic moment, the interfering polarized neutron beams are subjected to time-dependent magnetic fields, the phase shift having the expression $\delta = (1/\hbar)(\int_{path1} \mathbf{\mu} \cdot \mathbf{B}(t)dt - \int_{path2} \mathbf{\mu} \cdot \mathbf{B}(t)dt)$ where $\mathbf{\mu}$ is the magnetic

moment of the neutron. A phase space treatment of the scalar Aharonov-Bohm effect involves also the interference of two wavefunctions that differ through a phase shift, so that the approach presented in this paper is also applicable to this case.

FIGURES

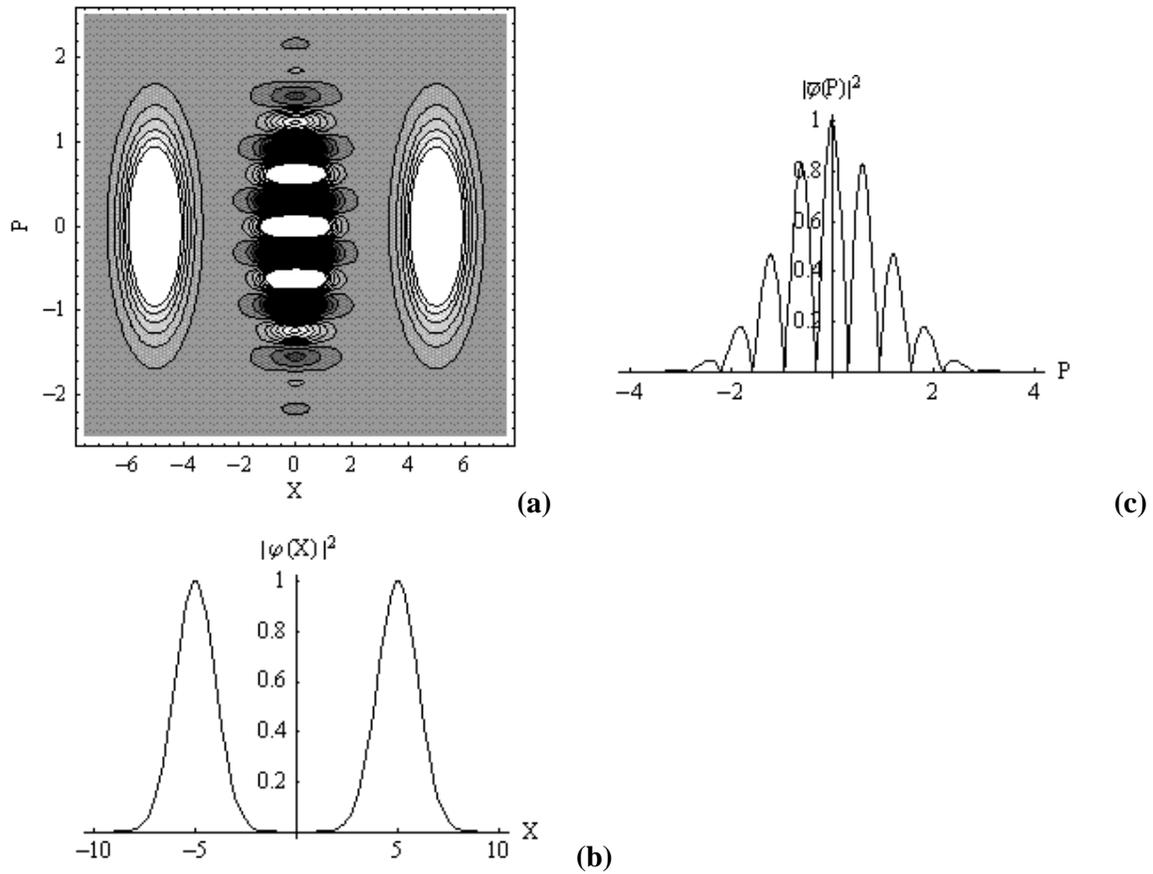

Fig.1 a) The WDF, b) $|\varphi(x)|^2$ and c) $|\bar{\varphi}(p)|^2$ for two Gaussian wavefunctions of width $x_0$ representing two interfering particle beams separated by $d = 5x_0$ immediately after the slits

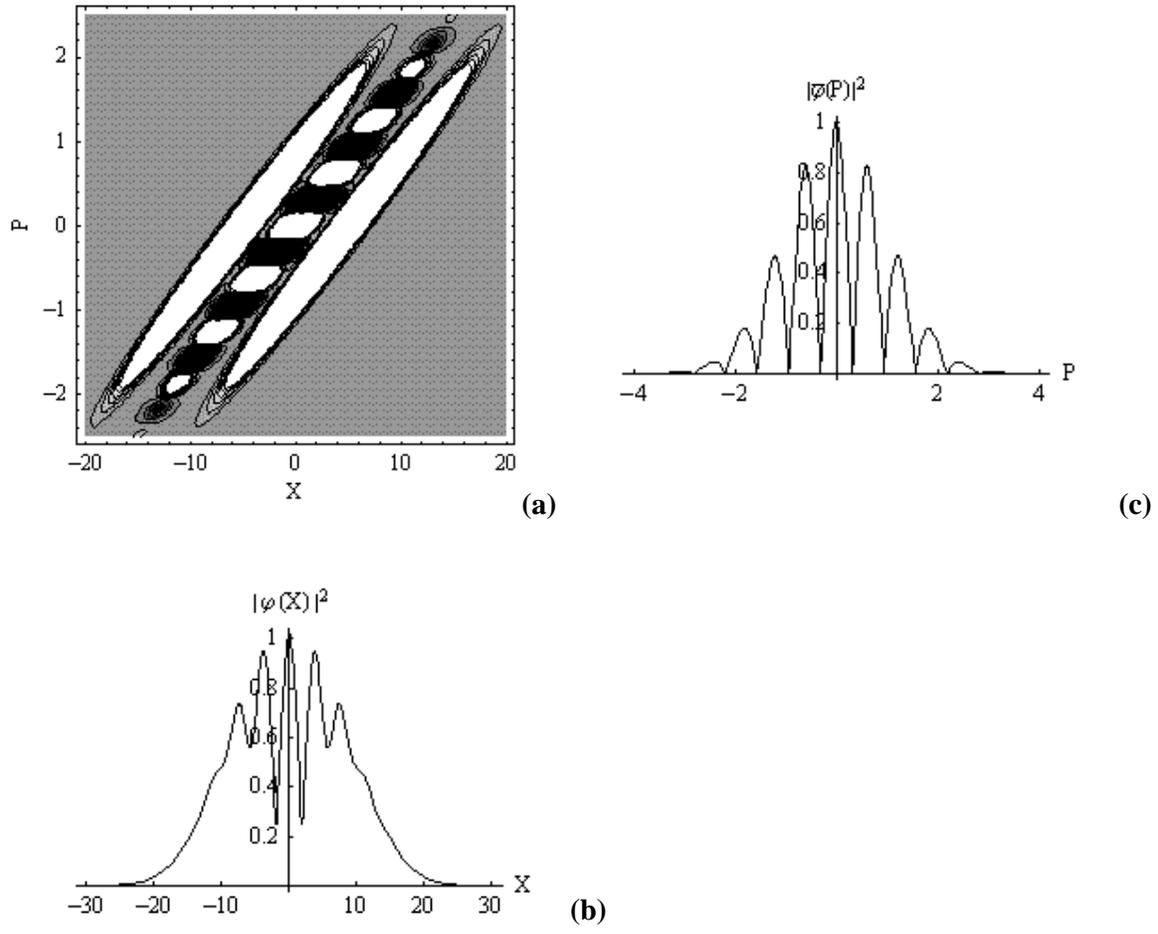

Fig.2 Same as in Fig.1 after propagation through a normalized distance $\alpha = 6x_0^2/\hbar$

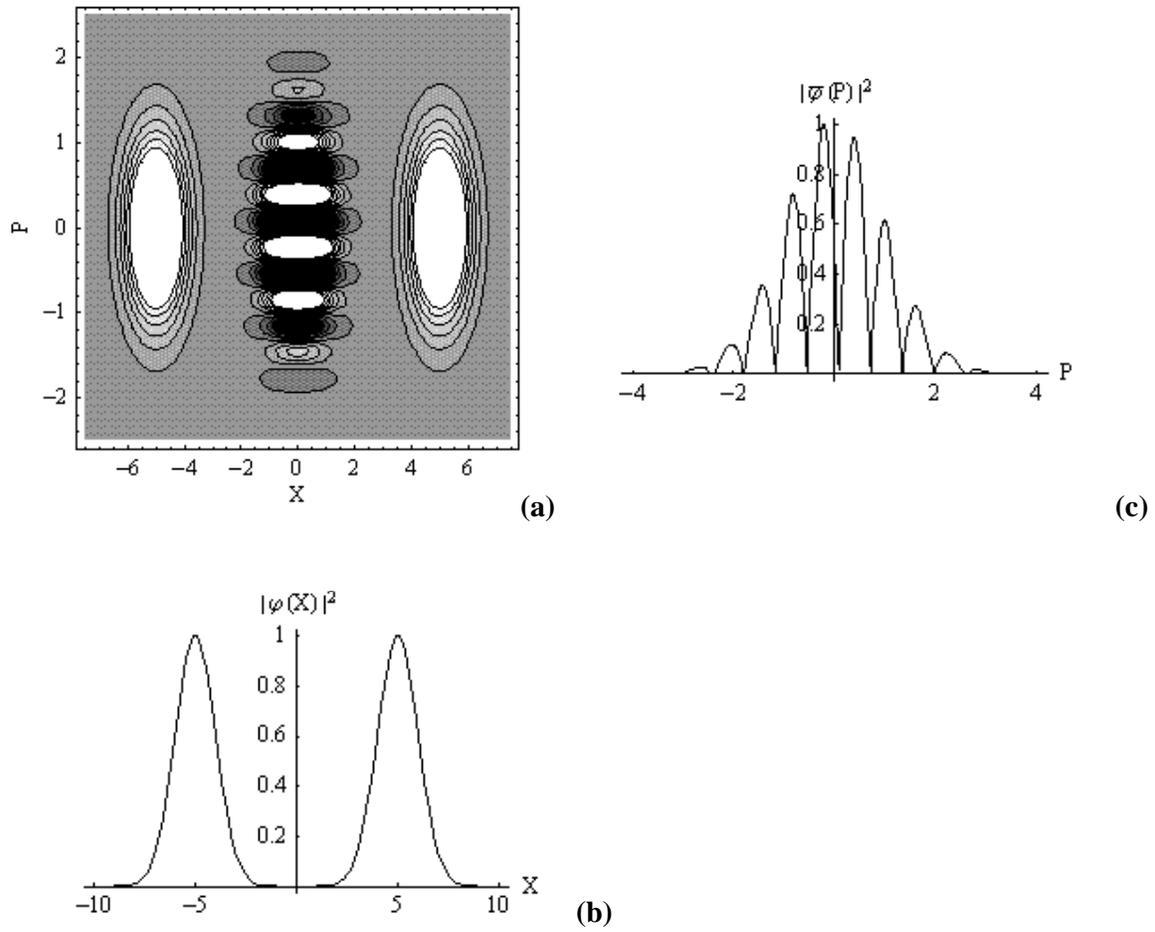

Fig.3  Same as in Fig.1 in the presence of the Aharonov-Bohm effect which induces a relative phase shift $\delta = 4$

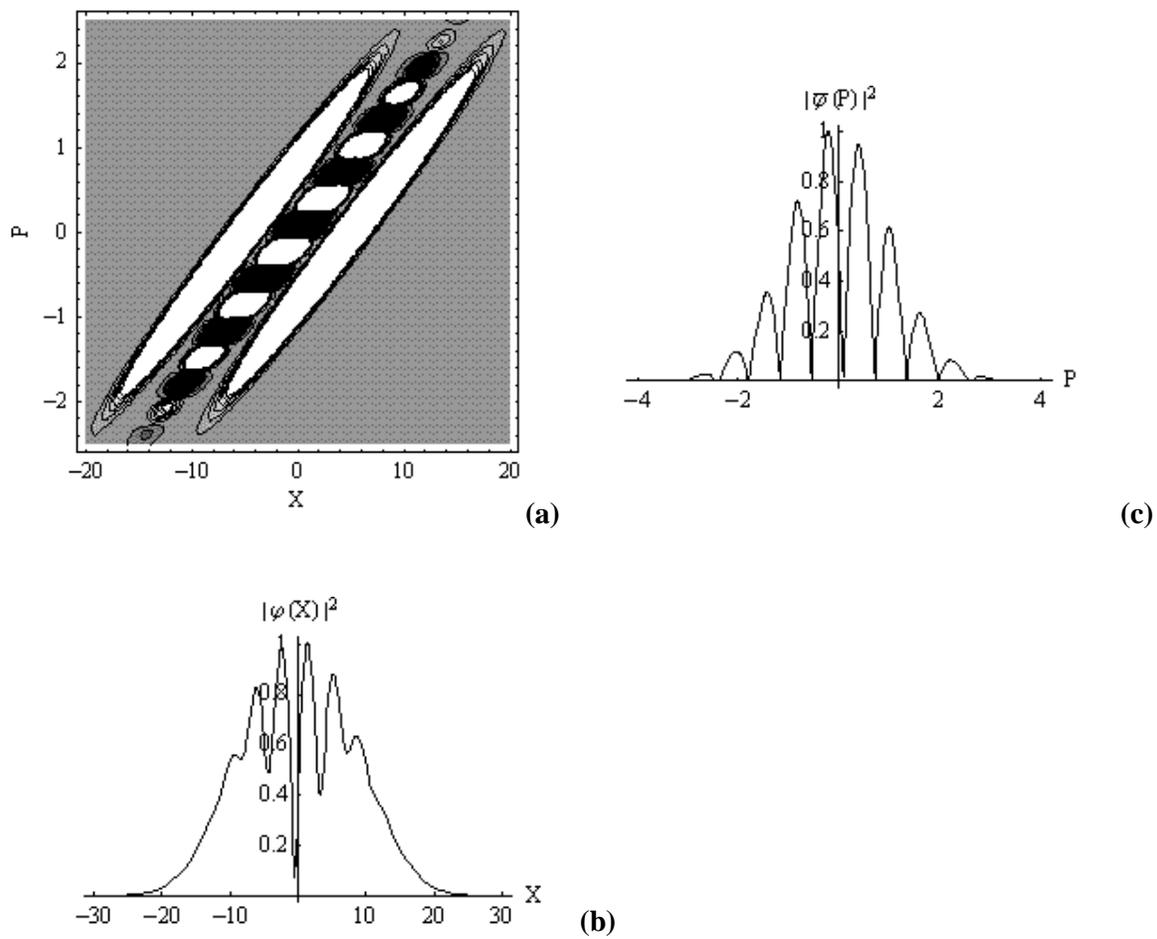

Fig.4   Same as in Fig.2 in the presence of the Aharonov-Bohm effect which induces a relative phase shift $\delta = 4$